\documentclass[%
 reprint,
 amsmath,amssymb,
 aps,
]{revtex4-2}

\usepackage{graphicx}% Include figure files
\usepackage{dcolumn}% Align table columns on decimal point
\usepackage{bm}% bold math

\usepackage{tcolorbox}
\usepackage{ragged2e}

\usepackage{multirow}%Added by AMALI
\usepackage{enumitem}

\begin{document}

\preprint{APS/123-QED}

\providecommand{\keywords}[1]
{
  \small	
  \textbf{\textit{Keywords---}} #1
}

\title{Assessing Scientific Practices in Physics Paper-based Assessments}

\author{Amali Priyanka Jambuge}
\email{amali@ksu.edu}
\affiliation{%
 Department of Physics, Kansas State University, Manhattan, Kansas 66506, USA
 }

\author{James T. Laverty}%
 \email{laverty@ksu.edu}
\affiliation{%
 Department of Physics, Kansas State University, Manhattan, Kansas 66506, USA
}

\date{\today}% It is always \today, today,

\begin{abstract}
Calls to transform introductory college physics courses to include scientific practices require assessments that can measure the extent to which these transformations are effective. Such assessments should be able to measure students' abilities to blend conceptually important concepts with practices scientists engage in. To design assessment tasks that can measure these abilities, we leveraged Evidence-Centered Design and the Three-Dimensional Learning Assessment Protocol with the focal scientific practice of “Using Mathematics." We conducted video recorded one-on-one think-aloud interviews to explore how students interpreted these tasks. In this paper, we articulate our design process and the analysis of students' responses using the ACER (Activation-Construction-Execution-Reflection) framework. Our assessment tasks elicited students' abilities to blend concepts with mathematics and written solutions elicited evidence of their abilities to blend them most of the time. This work extends our understanding on incorporating scientific practices into paper-based assessments at the introductory level. 

\end{abstract}

\keywords{Assessments, Introductory level, Scientific practices, Problem-solving}
%Use showkeys class option if keyword %display desired

\maketitle

\section{\label{Intro} Introduction}

There are recent calls to include scientific practices into college classrooms that underscore the importance of bringing student knowledge closer to its usage~\cite{Laverty2016,Cooper2015,Mcdonald2015,Singer2013,Kozminski2014}. This is in part to expose college students to the same learning environment as they have been exposed at K-12~\cite{Cooper2015}. As Cooper \emph{et al.}~\cite{Cooper2015} mention, “It would be a disservice to throw these students back into typical introductory courses…” Scientific practices constitute generalizable actions that scientists engage-in on a daily basis (such as develop and use models, analyze and interpret data, use mathematics, and plan and carry out investigation). Blending these practices with concepts core to physics (Core Ideas) promotes deeper learning~\cite{Framework2012,NGSS2013}. 

Incorporating scientific practices into college courses and evaluating the extent this transformation is effective requires assessments that have the ability to measure not just what students know, but how they use and apply their knowledge to new situations~\cite{Matz2018}. As Cooper~\cite{Cooper2016} mentions, “If we know what we are looking for, it is easier to recognize and assess it when we see it.” The K-12 framework for science education~\cite{Framework2012} well describes the scientific practices, allowing us to assess it~\cite{Cooper2016}. However, developing assessments that address scientific practices is identified as arduous and time consuming process~\cite{Cooper2015,Pellegrino2015,McElhaney2018,Osborne2014}. Thus, how can we assess students' work products (such as oral form of explanation or written solution) to articulate their abilities to engage in scientific practices along with concepts? 

In this work, we articulate a process for developing assessment tasks by focusing on the scientific practice of "Using Mathematics" and the concept of "force". Thus, our focus here is to assess how students use mathematics (hereinafter, math) to do physics rather than how they use math alone. Having a systematic, theory-driven approach to assess students' ability to use math would facilitate extending our understanding on assessing their ability to engage in the rest of the scientific practices as well.

We situate our study in interviews of students solving a paper-based exam simulating a summative assessment environment. The participants in this study are introductory-level students who weren’t specifically instructed with learning goals associated with scientific practices. 

In this paper, we build on work by Harris \emph{et al.}, and Stephenson \emph{et al.} to design assessment tasks that address scientific practices by leveraging principles of Evidence-Centered Design and to validate them for their potential to elicit expected evidence ~\cite{Harris2019,Harris2016,Mislevy2005,Stephenson2020,Mislevy2006,Mislevy2003,Mislevy1999}. The existing work on designing assessment tasks to assess scientific practices using Evidence-Centered Design covers middle school science students~\cite{McElhaney2018,Harris2019,Harris2016,Debarger2014,McElhaney2016,Pellegrino2015}, introductory-level chemistry students~\cite{Stephenson2020}, and upper-division physics students~\cite{Rainey2020,Jambuge2020}. We fill the gap in the literature by introducing a theory-driven methodology adopting Evidence-Centered Design to assess scientific practices in physics paper-based assessments at introductory-level. 

In the remainder of the paper, we try to answer our research questions:

\begin{enumerate}[noitemsep]
\item How do we develop assessment tasks to assess students' engagement in using mathematics along with physics concepts?
\item How can we validate students' work products in response to these tasks for their potential to elicit students' abilities to blend mathematics with concepts?
\item How much evidence of their abilities to blend mathematics with concepts do we get from looking at the written responses?
\end{enumerate}

% In the remainder of the paper, we try to answer our research questions, 1) How do we develop assessment tasks to assess students' engagement in using mathematics along with physics concepts?, 2) How can we validate students' work products in response to these tasks for their potential to elicit students' abilities to blend mathematics with concepts?, and 3) How much evidence of their abilities to blend mathematics with concepts do we get from looking at the written responses?

In Sec.~\ref{Literature}, we walk you through the existing literature on nature of the assessments in Physics Education Research (PER), assessment design and validation, and problem-solving and using math in physics. In Sec.~\ref{Theory}, we articulate the theoretical approach for our task design and validation process followed by research questions in Sec.~\ref{RQ}. In Sec.~\ref{Methods}, we explain our methodology for task design and the analysis of student responses to the designed tasks followed by data analysis exemplars. We finally provide some insights into our results suggesting potential implications for assessment design and validation along with future work (in Sec.\ref{Results and Discussion}, Sec.~\ref{Conclusion}).

\section{\label{Literature} Literature Review}

\subsection{Assessments in PER}

Assessments can be broadly viewed as either formative or summative. Formative assessments are used on a daily basis to support student learning by giving students feedback to reflect on their own learning and to adjust the subsequent instruction of the instructor. On the other hand, summative assessments are used to provide evidence of achievement to make decisions such as grading and retention~\cite{Pellegrino2014,NRC2001}. The available and widely used standardized assessments (such as concept inventories) in PER typically are used for summative purposes to gather evidence of what students have learned by the time they take them~\cite{Laverty2018}. As of now, there are almost 100 research-based assessments available for the physics education community as listed in the PhysPort website~\cite{PhysPort}. These assessments are identified by PhysPort as assessing content knowledge, problem-solving, scientific reasoning, lab skills, beliefs/attitudes, and interactive teaching.  

These standardized assessments primarily measure students’ conceptual knowledge (63 out of 95 assessments measure content knowledge) in numerous physics concepts~\cite{PhysPort}. Thus, these off-the-shelf assessments have a significant impact on education reform by providing a universal way of evaluating student understanding that leads teachers to assess and revise their teaching methods~\cite{Madsen2017}. For example, these assessments have been used to evaluate teaching methods~\cite{VonKorff2016,Freeman2014,Hake1998}, learning outcomes of different student populations~\cite{Madsen2013,Lorenzo2006,Brewe2010}, and curriculum reforms~\cite{Caballero2012,Kohlmyer2009}. 

The most common standardized assessments used at introductory level are Force Concept Inventory~\cite{Hestenes1992}, Force and Motion Conceptual Evaluation~\cite{Thornton1998}, Brief Electricity and Magnetism Assessment~\cite{Ding2006}, and Conceptual Survey of Electricity and Magnetism~\cite{Maloney2001}. While these concept inventories are assets in eliciting students’ conceptual understanding, they are not designed to elicit students’ engagement in scientific practices~\cite{Laverty2018}. However, calls to include scientific practices into K-12 level and college curricula brought assessment developers’ attention to design tasks to assess students’ abilities to engage in scientific practices and concepts. For example, Wolf \emph{et al.}~\cite{Wolf2019} developed and validated a practical exam to assess student abilities to engage in scientific practices in introductory physics laboratories. While this work provides a promising way to assess scientific practices in laboratory settings, it is unclear how this approach can be generalizable to typical large-scale introductory classrooms where paper-based assessments play a prominent role. 

\subsection{Assessment Design and Validation}

Assessments give us vital information about students' learning. The ``information'' refers to the types of inferences we make out of students' work, attributing a certain set of knowledge and skills to the student performance that align with the designer’s goal for the assessment. The process of making inferences is referred to as ``reasoning from evidenc'' that describes the process of drawing inferences accumulating a set of supporting evidence from students' work~\cite{NRC2001}.

This process can also be portrayed as a triangle where the triads represent the three key elements highlighted in designing assessments, the \emph{assessment triangle}, a model of student cognition and learning in the domain, a set of beliefs about the kinds of observations that will provide evidence of student competencies, and an interpretation process for making sense of the evidence~\cite{NRC2001}. The assessment triangle conceptualizes the nature of assessment tasks, but an elaborative framework is needed to operationalize those conceptualizations. 

Evidence-Centered Design is embedded in the logic of the assessment triangle. It provides a methodological and systematic approach to the assessment task design that helps elicit students' proficiencies attributed to the designer’s intention. It has also been identified as a promising approach for developing assessment tasks that effectively measure concepts intertwined with scientific practices~\cite{Pellegrino2014,NRC2001}. In particular, there are several work, where researchers have adopted Evidence-Centered Design to design assessment task that assess scientific practices and concepts~\cite{Harris2019,Harris2016,Stephenson2020,McElhaney2018,McElhaney2016,Pellegrino2015,Debarger2014}.

Assessment task design is accompanied by validating the designed tasks. There are several approaches to task validation in the research literature. One approach takes the form of content validity where the alignment between the task content with the subject matter framework is evaluated by experts in a particular domain~\cite{NRC2001}. Extending this approach to include empirical evidence to determine the extent to which designed tasks tap the intended cognitive processes is also emphasized in several work~\cite{NRC2001,Messick1993}. 

Thus, the argument-based approach to validity consists of two parts: interpretive and validity arguments~\cite{Kane2006,Kane2013}. First, the interpretation and use of assessment scores are proposed prior to disseminating the assessment tasks to students (interpretive-argument). Second, the plausibility of the interpretive-argument is validated via student think-aloud interviews  (validity-argument)~\cite{NRC2001,Pellegrino2014,Adams2011}. During this process of validation, students' unintended problem-solving approaches that tap unintended cognitive processes differing from the designer's intention can be documented. Thereby, the iterative modifications to task design can be made until the proposed interpretations and use of assessment scores are reasonable. Evidence-Centered Design, in particular, leverages the argument-based validity approach to validate assessment tasks where the claims about student knowledge and skills are backed by evidence~\cite{Bennett2017,Gane2018}. We acknowledge that there are numerous approaches for task validation in the research literature other than the approaches we described in this section (e.g., criterion validity, classical test theory, item response theory). Such approaches are not the focus of this study thus, we don't provide an extensive literature review on that.

\subsection{Problem-solving and "Using Math" in Physics}

Mathematics is one of the cornerstones in physics problem-solving. However, use of mathematics in physics is found to be different from mathematics alone~\cite{Redish2006,Uhden2012,Lopez2015,Michelsen2006,Karam2014}. This nuance often causes problems due to the gap between student and instructor expectations of what it means to do math in physics. 

Physicists believe use of math in physics occurs in a certain, prescribed way~\cite{Uhden2012,Redish2006}. Thus, one way to evaluate students’ use of math is to probe their work products produced during problem-solving with the prescribed models for using math in physics. For example, Redish~\cite{Redish2006} developed a model describing the bare bones of how to use math in physics. This model includes 1) Mapping the physical system into a mathematical model, 2) Processing the mathematical model to simplify it, 3) Interpreting the results obtained to explore what they tell about the physical system, and 4) Evaluating the result to validate its extent to accurately represent the physical system. 

However, it is worth noticing that students do not necessarily follow that clean-cut procedure when solving problems~\cite{Redish2006}, else they approach problems in ineffective ways~\cite{Reif1976}. These ineffective ways might arise due to the lack of a systematic strategy that guides students to apply their knowledge. Thus, lots of research work has targeted teaching students specific problem-solving strategies~\cite{Reif1976,Wright1986,Heller1992} and to evaluate students' engagement in problem-solving~\cite{Heller1992,Murthy2007,Docktor2009,Yerushalmi2012,Foster2000,Mason2010,Docktor2016}.

Another approach to evaluate students’ use of mathematics is to explore how students use math on their own terms. The body of research work on this aspect leverages theoretical perspectives such as resources~\cite{Hammer2000}, framing~\cite{Hammer2005}, and epistemic games~\cite{Collins1993} to explore students’ in-the-moment reasoning while solving physics problems~\cite{Caballero2015}. Resources are the fragments of knowledge being activated based on how students tacitly determine what kind of knowledge might be appropriate for the problem at hand (framing). This leads to a set of locally coherent activities ("moves") students do during problem-solving (epistemic games). 

One such study that leverages the aforementioned theoretical perspectives is the work by Tuminaro and Redish~\cite{Tuminaro2007} where they observed six epistemic games introductory students play while solving physics problems. The tacit judgement students make to decide which game to play depends on their expectations for the problem at hand. These expectations determine which resources to bring into a particular problem context. Bing and Redish~\cite{Bing2009} leveraged resources and epistemological framing to capture how upper-level students use math. A recent study by Modir, Thompson, and Sayre~\cite{Modir2017} developed a theoretical framework that models upper-level student framing in math and physics adapting epistemological framing.  

While theoretically well-grounded approaches are more robust than prescribed models to explore student use of math in physics, they pose challenges on instructors who are not familiar with these theoretical constructs. Attending to these challenges, the ACER framework~\cite{Wilcox2013,Caballero2015} bridges the gap between prescribed models for student use of math with the resources framework and epistemic frames. 

ACER stands for the Activation of the tool, Construction of the model, Execution of the mathematics, and Reflection of the results. These components are pertaining to the activation of the mathematical tool, mapping between the physics and mathematics of a problem, working with the procedural aspects of the mathematical tools, and interpreting and checking the intermediate and the final steps of the solutions respectively. Each of these components (epistemic frames) consist of several subcodes (resources) in which students shift back and forth while solving physics problems. These subcodes are not categorized in any specific order, rather it describes what steps students might take while going through the problem. For example, in construction of the model component, students might be making assumptions or developing a representation that describes the physical system. The subcodes under the components rely on the nature of the assessment tasks.

\section{\label{Theory}Theoretical Approach}

In this section, we articulate the theoretical approach to our task design process adapting Evidence-Centered Design and the Three-Dimensional Learning Assessment Protocol~\cite{Laverty2016} along with our theoretical assumptions for task validation. We first articulate the general principles of Evidence-Centered Design as laid out by its developers and then how researchers adapt that to incorporate scientific practices. We then explain the utility of the Three-Dimensional Learning Assessment Protocol into our work. We also provide insights into our task validation approach within the Evidence-Centered Design. 

\subsection{Evidence-Centered Design}

Employing educational assessments can be viewed as a process of reasoning from evidence, i.e., how we can use assessments to infer what students know and can do~\cite{NRC2001}. However, designing assessments to measure these constructs requires careful and thoughtful approaches. As Mislevy~\cite{Mislevy2005} mentions, “Assessment design is often identified with the nuts and bolts of authoring tasks. However, it is more fruitful to view the process as first crafting an assessment argument, then embodying it in the machinery of tasks…” This way the distinction between testing and assessment is emphasized.  

Drawing from previous work, beyond this point, we explain the basics behind Evidence-Centered Design (ECD)~\cite{Mislevy2005, Mislevy2006, Mislevy1999, Mislevy2003}. ECD suggests that we first gather substantial information of the domain of interest (such as physics). This substantial information includes, but is not limited to, concepts, student knowledge representations, and terminologies. Then the information gathered can be depicted into a design pattern. 

Design pattern comprises several elements to ensure coherent nature between the claim about what students should know and be able to do (Student Model), expected evidence to meet the claim (Evidence Model), and the task to elicit the evidence (Task Model). The Student Model articulates the knowledge, skills, and abilities identified as important. The evidence for these knowledge, skills, and abilities are required to justify the claim about what students should know and be able to do. The Evidence Model articulates the potential observations in the student work that constitute evidence for knowledge, skills, and abilities. The "Task Model" makes sure that the task features have the potential to elicit potential observations in students' work. 

After laying out the basics of ECD, we now turn to work that utilizes ECD as a design approach to design assessment tasks that assess scientific practices articulated in the framework for K-12 science education~\cite{Framework2012}. The theoretical views below mostly capture the ideas in Harris \emph{et al.}~\cite{Harris2019}, and we suggest this reference for readers who are interested in the detailed assessment task design approach laid out here. 

Our assessment task design approach, which is mostly reflected the approach by Harris \emph{et al.}~\cite{Harris2019} is also built around the three models, i.e., student, evidence, and task models. The Student Model, claims about what students should know and be able to do takes the form of learning performances. Learning performances articulate assessable statements that measure student abilities to blend scientific practices with concepts. The knowledge, skills, and abilities required to meet the learning performances are also articulated in the Student Model. Evidence Model consists of evidence statements that provide evidence that students have the required knowledge, skills, and abilities. The Task Model makes sure that the assessment tasks have the potential to elicit the evidence statements. 

As we develop tasks for the introductory-level physics students, it is worthwhile to explore the valued scientific practices and the ways those can be elicited in assessment tasks in introductory level. Thus, we next bring your attention to the Three-Dimensional Learning Assessment Protocol (3D-LAP)~\cite{Laverty2016}, a tool that can be used to design assessment tasks to elicit student abilities to engage in scientific practices. The 3D-LAP consists of criteria each for scientific practice, and to align with a scientific practice, all of the underlined criteria should be met. This criteria was developed with a team of disciplinary experts that consisted of researchers in the field of education-based research and more traditional faculty members. The 3D-LAP was successfully validated for its reliability to differentiate tasks that have the potential to elicit scientific practices and concepts with the tasks that do not have the potential to do so~\cite{Laverty2016}.

To have a coherent task design, we couple the 3D-LAP with the ECD. In other words, the criteria in the 3D-LAP for tasks to elicit scientific practices can be used as task features in the Task Model in ECD which we explain with more details in the Sec.~\ref{Methods}. 

\subsection{Assessment Task Validation}

Assessment task design is followed by the validation of those tasks~\cite{Harris2019,Stephenson2020}. Adapting the 3D-LAP, a tool that has been validated for its reliability to differentiate assessment tasks with and without having potential to elicit scientific practices along with concepts, contributes to our tasks' content validity~\cite{Laverty2016}. The assessment task validation also ensures the extent students demonstrate the evidence that the tasks intended them to be showcased. One way to evaluate such validity is to examine the processes students go through when they encounter these tasks and look for evidence to determine that the task functions as intended. In this way, the assessment tasks can be connected with the students’ ideas~\cite{Madsen2017}. In particular, the student solutions should be explored in light of evidentiary arguments to determine the extent to which assessment tasks have the potential to elicit appropriate predefined evidence. Think-Aloud~\cite{Ericsson1980,Fonteyn1993} interviews have been suggested as a way of eliciting student problem-solving processes to the assessment tasks~\cite{NRC2001,Pellegrino2014,Adams2011}.

The task validation process requires us to allow descriptive, unexpected student evidentiary representations to take into consideration. In other words, the predefined evidence statements that give us the evidence that students have targeted knowledge, skills, and abilities can be modified based on student solutions to entirely capture their potential to elicit the learning performance. As part of these modifications, students' fine-grained evidentiary representations that pertain to the evidence statements can emerge. Thus, an analytic framework that closely captures the predefined evidence statements can be adapted to interpret student work products. 

We expand our ESs, and thereby the “Evidence Model” by coupling with an analytic framework that closely captures the predefined ESs which is the ACER framework in our case. %Thus, each ES in Table~\ref{Table1} is aligned with a component of the ACER framework respectively. 
We adopted the ACER framework because 1) the component in the framework well-aligned with our predefined ESs for assessment tasks, 2) it gives insight into learning theories~\cite{Caballero2015} while remaining open for instructors who are not familiar with the theoretical constructs of the framework, and 3) its emphasis on organizing students' written work products (as compared to only video data). This approach modifies the “Evidence Model” by introducing student knowledge representations based on their own terms.

% \begin{figure}[t]
%   \includegraphics[width=\linewidth]{Fig1.jpg}
%   \caption{Car accident reconstruction problem from the assessment.\label{Fig1}}
% \end{figure}

\begin{table*}[htb]
\caption{Stages included from the ECD process for the task in Fig.~\ref{Fig1}. Task features are the criteria in the 3D-LAP to design constructed response assessment tasks to elicit the learning performance associated with the scientific practice of “Using Math” blended with ``force''. The * represents the elements added as part of the task’s validation process based on student responses.}
\centering
\begin{tabular}{p{2cm}|p{15.5cm}}
\hline\hline
Learning Performance & Students will be able to use math to determine kinematic values from data about the motion presented and use that information to reach a conclusion about the nature of the motion. \\
\hline
Knowledge, Skills, and Abilities & \begin{tabular}[c]{@{}l@{}}KSA1: Identify kinematics principles as appropriate to determine the nature of the motion. \\ KSA2: Identify relevant physics equations or generate mathematical equations to connect the variables in the \\ physical system. \\ KSA3*: Conduct appropriate mathematical manipulations.\\ KSA4: Determine the nature of the motion.\end{tabular} \\
\hline
Evidence Statements & \begin{tabular}[c]{@{}l@{}}ES1: Statements of the unpacking of appropriate physics concepts to solve the problem. \\ ES2: Statements of the use of mathematical equations that represent the given physical system.\\ ES3*: Statements correspond to mathematical manipulations.\\ ES4: Statements interpreting the results from the mathematical manipulations.\end{tabular} \\
\hline
Task Features & \begin{tabular}[c]{@{}l@{}}Question gives an event, observation, or phenomenon. \\ Question asks students to perform a calculation or statistical test, generate a mathematical representation, \\ or demonstrate a relationship between parameters. \\ Question asks students to give a consequence or an interpretation (not a restatement) in words, \\ diagrams, symbols, or graphs of their results in the context of the given event, observation, or phenomenon.\end{tabular} \\
\hline\hline                                        
\end{tabular}
\label{Table1}
\end{table*}

\section{\label{RQ}Research Questions}

Our research questions articulated in Sec.~\ref{Intro} turned in to a form below after incorporating the theoretical perspectives we lay out in Sec.~\ref{Theory}. Thus, in this work, we try to answer the research questions, 

\begin{enumerate}
  \item How do we develop assessment tasks to assess student engagement in learning performances that blend scientific practices and concepts?
  \item How can we validate student work products in response to these tasks for their potential to elicit expected evidence to achieve the target learning performances?, and
  \item How much evidence of their abilities to meet the learning performances do we get from looking at the students' written responses? 
\end{enumerate}

\section{\label{Methods}Methodology}

As we move forward on this section, we explicate our assessment task design process, data collection, and data analysis to answer our research questions in Sec.~\ref{RQ}. The presented methodology in this section does not reflect the exact process we followed during our research. We modified and optimized the process based on our research experience.  

\subsection{Assessment Task Design}

Harris \emph{et al.}~\cite{Harris2019} articulated their task design process adapting ECD along with multiple design stages to ensure coherent task design to intertwine concepts with practices. We build on that work to design assessment tasks in the context of undergraduate physics, specifically introductory mechanics. Table~\ref{Table1} summarizes the stages in the ECD process (described below) used to develop the task shown in Fig.~\ref{Fig1}.

\begin{figure}[b]
    \centering
    \begin{tcolorbox}[colback=white,arc=0mm,colframe=black]
      \justify{Assume you are responsible to carry out an accident reconstruction case at your local police station.  The car accident left a skid mark of length $40.3~m$ on the road.  The driver claims he was driving under the speed limit.}
      
      \justify{In order to further clarify this case, you did an experiment at a crash site with similar accident conditions.  The data shows an average skid mark of length $22.4~m$ when the brake was locked while the car was travelling at the speed of $15.2~m/s$.}
      
      \justify{Describe how you can determine the speed of the car before the accident.}
      
      \justify{Your job is to determine whether or not the driver was speeding before the car accident. If the speed limit of the area that the accident occurred in is $18~m/s$, is the driver at fault?}
    \end{tcolorbox}
    \caption{Car accident reconstruction problem from the assessment.}
    \label{Fig1}
\end{figure}

We first need to identify what we value that students should know and be able to do in the domain of physics. We then construct an assessable statement that blends what students should know (concept) and be able to do with their knowledge (scientific practice) in the form of a Learning Performance (LP). We then determine the Knowledge, Skills, and Abilities (KSAs) to achieve that LP. Then we articulate the Evidence Statements (ESs), which specify what we need to see in a student's response to demonstrate that they have the KSAs we articulated previously. In the final stage, we define the task features needed to elicit the evidence articulated in the ESs. 

As we stated in Sec.~\ref{Theory}, the criteria in the 3D-LAP lays out the basis for the task features to elicit the ESs~\cite{Laverty2016}. The protocol consists of a set of criteria for each scientific practice where all the specifications of the criteria should be satisfied in order for an assessment task to have the potential to elicit a scientific practice. For example, to elicit the scientific practice of using mathematics, we should develop the task such that it 1) gives an event, observation, or phenomenon 2) asks students to perform a calculation or statistical test, generate a mathematical representation, or demonstrate a relationship between parameters, and 3) asks student to give a consequence or an interpretation (not a restatement) in words, diagrams, symbols, or graphs of their results in the context of the given event, observation, or phenomenon. The phenomenon can be integrated with concepts around core ideas in physics (in our case, force) to take the form of task features to elicit a LP that addresses the scientific practice of “Using Math”.

%For example, by including ``Assume you are responsible to carry out an accident reconstruction case at your local police station" in the assessment task in Fig.~\ref{Fig1}, it met the task feature 1 in Table~\ref{Table1} by providing an event to consider. To meet the task feature 2, assessment task includes ``Describe how you can determine the speed of the car before the accident", given the data from the simulated crash site. By including ``If the speed limit of the area that the accident occurred in is 18 m/s, is the driver at fault?" in the assessment task, it met the task feature 3 that requires an interpretation for students' answers.     

Similarly, each task of the assessment is accompanied by a logical argument that can be built by following the aforementioned generalized procedure. The authors of this manuscript discussed and refined the assessment tasks until each one met all the criteria in the 3D-LAP (task features) to elicit student abilities to engage in the scientific practice of “Using Math” blended with ``force''.  

\subsection{Data Collection}

We conducted Think-Aloud interviews~\cite{Ericsson1980,Fonteyn1993} with students to answer our second and third research questions. Think-Aloud protocols have been used with individuals with varying levels of expertise in a domain of interest to articulate the information these individuals attend to at a given time and how this information is organized during problem-solving~\cite{Fonteyn1993}. Interviewers ask subjects to ``think-aloud'' and verbalize their thought processes while performing cognitively demanding tasks such as problem-solving. According to Ericsson and Simon~\cite{Ericsson1980}, a subject's verbalization that occurs simultaneously during problem-solving does not alter their thought processes as long as the interviewer does not interrupt with probes. 

The participants of our study were students in first or second semester introductory-level, calculus-based physics courses. The students voluntarily participated in this study, and they were remunerated with twenty dollars in gift cards for their participation. We scheduled individual interview sessions that facilitated a quiet environment for subjects to think-aloud simulating an exam environment~\cite{Fonteyn1993}. We asked students to think-aloud while working on the assessment tasks. For each student, the think-aloud interview lasted about one hour. Similar to an exam, and in keeping with the think aloud protocol, the interviewer did not assist the students with the problems or answer questions about the problems. Like a normal exam, in our interviews, students moved back and forth between problems as they wished, and they determined when they were done with each particular problem.  

When students paused for several seconds, the interviewer reminded them to keep thinking aloud. We took notes during the interview that can be followed-up when the interviewee finished the tasks. This led us to further clarify subjects' problem-solving processes and reasoning. The interviews were video and audio recorded and work products in the form of written solutions were collected and scanned for further analysis. 

We did not include interviews with audio issues, and interviews where students did not regularly think-out aloud even after being encouraged to do so several times by the interviewer. Overall, we had 7 distinct assessment tasks that addressed the scientific practice of “Using Math” among 8 interviews, thus giving us 56 total instances for the analysis. Out of the 56 instances, in 3 instances students did not respond to the assessment tasks. Thus, we transcribed remaining 53 student verbal responses both manually and using an AI transcription service~\cite{Otter}. We corrected some of the transcriptions obtained from AI transcription service for their clarity. The accompanying 53 written solutions were gathered for the analysis. All names used in this manuscript are pseudonyms. 

\begin{table*}[htb]
\caption{Portion of the codebook used to analyze data. Each subcode is assigned with a symbol to make the navigation in between subcodes efficiently. See Appendix A for the full codebook with definitions and examples.}
\centering
\begin{tabular}{p{4cm}|p{1.5cm}|p{12cm}}
\hline\hline
Component & Subcode & Description of the Subcode \\
\hline
\multirow{2}{*}{Activation (A) $\sim$ES1} & A1 & Identify appropriate physics concepts. \\
 & A2 & Identify general physics equations to be applied. \\
 & A3 & Identify target parameters. \\
\hline                                            
\multirow{2}{*}{Construction (C) $\sim$ES2} & C1 & Apply the general equations to a particular situation. \\
 & C2 & Make assumptions. \\
 & C3 & Develop representations. \\
 & C4 & Develop mathematical relations based on the concepts used. \\
\hline                                            
\multirow{2}{*}{Execution (E) $\sim$ES3} & E1 & Manipulate symbols.\\
 & E2 & Perform an arithmetic calculation. \\
 & E3 & Execute math conceptually. \\
 & E4 & Substitute expressions. \\
 & E5 & Manipulate mathematical expressions. \\
\hline                                            
\multirow{2}{*}{Reflection (R) $\sim$ES4} & R1 & Make sense of the answer with the information given in the prompt. \\
 & R2 & Make sense of the answer found in an intermediate/final step. \\
 & R3 & Make sense of the result for use in a subsequent step. \\
\hline\hline
\end{tabular}
\label{Table2}
\end{table*}

\subsection{\label{Analysis}Data Analysis}

In this section, we provide our data analysis approach that helped us answer our second and third research questions. We first provide insights into how we developed our codebook using the ACER framework to analyze student data, incorporating their own knowledge representations~\cite{Wilcox2015a,Wilcox2015b,Weliweriya2016}. We then demonstrate how we code verbal and written responses, followed by inter-rater reliability of the coding.  

\subsubsection{Code Book}

%Our analysis began with developing a codebook (See Table~\ref{Table2} for a portion of the codebook) that well represented student work. By doing so, we expanded our understanding of the ESs in Table~\ref{Table1} capturing multiple knowledge representations that give evidence for required KSAs in a way that they present in student work products. This approach addresses the validity aspects of our assessment tasks by articulating the ways students interpret these assessment tasks on their own terms. Moreover, the analysis coupled with the ACER framework gives us insights into the extent these tasks elicited the expected evidence to infer students' abilities on mathematical reasoning in the context of physics.   

To develop the codebook (see Table~\ref{Table2}), we started by selecting one assessment task and going through all students' responses to that task before looking at another task. %Then we iteratively searched students' transcribed verbal responses along with the written solutions to develop subcodes aligning with each component in the ACER framework. 
We carried out the coding process looking for appropriate subcodes, merging them when they overlapped. We finalized our codebook when no additional subcodes were identified as needed to represent students' work products, i.e, the codebook was saturated. The codebook also captures errors students make while solving the problems by including an “X” in the code. 

\subsubsection{\label{coding verbal and written}Coding Verbal and Written Responses}

We now turn to the goal of identifying if the tasks were capable of eliciting evidence to achieve the LP. If the assessment tasks have the potential to elicit the expected evidence, they should provide evidence for each component in the ACER framework (that is, activation, construction, execution, and reflection). 

Once the codebook was finalized, we coded the students' transcribed verbal responses sentence by sentence.  %(See Fig.~\ref{Fig2}, for example) 
Once a student's verbal response to an assessment task was coded, we compiled the subcodes corresponding to that problem-solving into a list (the ``verbal-codes''). %The subcode structures signal the processes that students went through while solving the problem. The similar coding process has been carried out for all the 53 students' transcribed verbal responses. 
%The student written solutions include, but are not limited to written assumptions/simplifications of the physical system considered, mathematical expressions, series of lines corresponding to the mathematical manipulations, and written interpretations made using the resultant mathematical expressions. Such features of
Then, the written solutions were coded by assigning an appropriate subcode to each line of a student's solution. %(See Fig.~\ref{Fig3}, for example). 
Once a student's written solution for an assessment task was completely coded, we compiled the set of subcodes corresponding to that problem solution into another list (the ``written-codes'').
The student's verbal and written codes were then synthesized into a single coding pattern (the ``combined-codes'') which constitutes their overall problem-solving approach. The motivation to obtain combined-codes is to capture a student's complete problem-solving approach. This process was repeated for all 53 verbal and written student responses.

For each assessment task, we analyzed the combined-codes across all students in our data set to explore the task's potential to elicit the expected evidence. This gives us evidence whether or not the intended cognitive processes were tapped during students' problem-solving. If the task was able to elicit the expected evidence -- at least one code from each ACER component; A, C, E, and R -- from a majority of the students, we determined that the task was good enough to differentiate student abilities to meet the LP. We define majority in our context as $>50\%$ of students.

If the assessment tasks elicited the expected evidence, we further explored the extent to which the students' written solutions (which are what typically get graded in coursework) accurately reflected their overall engagement with the problem. In order to determine if students' written solutions provided enough evidence to support the claim that they were or were not achieving the LP, we compared each students' written-codes with their combined-codes.

On the other hand, if the task was not eliciting the expected evidence, the component(s) of the ACER which is lacking was documented. This is for the future revisions of that task to deliberately elicit that component(s).

%In Sec.~\ref{Analysis}, we demonstrate detailed example for this generalized coding procedure for both verbal and written responses.

%Paul's attempt to rearrange the Fig. 2 and Fig.3 together.

%\begin{figure}[t]
%  \centering
%  \begin{subfigure}{0.5\textwidth}
%    \includegraphics[width=\linewidth]{Fig2.jpg}
%    \caption{Portion of the coded verbal response.}
%    \label{Fig2}
%  \end{subfigure}
  %
%  \begin{subfigure}{0.5\textwidth}
%    \includegraphics[width=\linewidth]{Fig3.jpg}
%    \caption{Portion of coded written response.}
%    \label{Fig3}
% \end{subfigure}
%  \caption{Example showing a portion of coded student verbal~(a) and written~(b) responses. The sub codes in verbal include A3 (green), A2 (green), %C1 (red), E2 (yellow), and R1 (blue). See Table~\ref{Table2} for detailed explanations of these sub codes.}
%  \label{Fig2comb}
%\end{figure}

%Paul's attempt end here.

\subsubsection{Example Coding}

Catherine started the accident reconstruction problem (see Fig.~\ref{Fig1}) by going through the problem statement and trying to make sense of it. Then, she referred to the equation sheet looking for information that she can relate to the problem. Catherine vocalized her initial thoughts on the problem as follows,

\begin{quote}
\emph{“...the car is going from a, hmm, faster speed down to a stop I can assume that the, hmm, initial velocity is equal to the, hmm, speed you travel at [inaudible] final velocity is equal to zero just until he stopped and since I have the length of skid mark I can assume that the initial position was zero and then the final position is how long that skid mark was.”}
\end{quote}

She also realized that she was not given any information related to time. Therefore, she chose to use the general equation \( V^2 = V_0^2 + 2a(x-x_0) \) to solve for acceleration. 

\begin{quote}
\emph{“Hmm, I do not have the time for any of those states. So hmm I... am so this is saying that if the brakes were locked so hmm that's kind of the maximum hmm decrease in acceleration hmm so I'm going to use hmm V squared equals V naught squared plus two a in parenthesis x minus x note hmm [A2]. That way I can solve for a [A3] ...”}
\end{quote}

Thereafter, she made an assumption that if she knows the acceleration at the crash site, the same acceleration can be used at the real accident. With this assumption in mind, she applied the general equation activated to the crash site to figure out the value for the acceleration. She manipulated the numerical value for the acceleration using the calculator. 

\begin{quote}
\emph{“because assuming that the actual accident, the driver locked the brake then to then and I'm just using that same acceleration to see if the driver was at fault or not [C2]. So hmm having zero squared equals fifteen point two meters per second squared plus two a and then in parenthesis it is the twenty two point four meters [C1]. So then just solving for a, [inserting values in the calculator to find the numerical value for acceleration, a] [E2].”}
\end{quote}

She reflected on the negative value obtained as the acceleration to make sense that it was a reasonable answer as the driver was going from a faster speed down to a lower speed by stating,

\begin{quote}
\emph{“The acceleration is equal to negative five point two one six which once again is a reasonable answer since they are going from a faster speed down to a lower speed. [R2]”}
\end{quote}

Her goal for this problem was to see when she applied the same acceleration to the actual accident to see what skid mark length it would give and then to compare it with the skid mark given for the actual accident. Thus, it is not that she came up with the answer, but reflected on the answer to see how she can use that information in subsequent steps of the problem-solving.

\begin{quote}
\emph{“Hmm then I'm going to take that acceleration and plugging into the exact same problem [R3] hmm to see if hmm the skid mark length that I get is equal to the actual skid mark length. [A3]”}
\end{quote}

Then she applied the general equation activated before to find the skid mark length given the speed limit 18 m/s and calculated the numerical answer using a calculator. 

\begin{quote}
\emph{“Well, I know that x note is gonna be zero just as I'm going from no skid mark to the length I'm just solving for the x [A3], so it will be V is equal to zero again and then I'm going to use V eighteen per second squared two times the negative five point meters per second and then in parenthesis it's x since x minus zero is just x [A2][C1]. Hmm so solving for that is [Calculating the numerical value for x using the calculator] [E2].”}
\end{quote}

Once she got the value for the skid mark length as thirty-one point zero five eight meters, she made a comparison with the given value of forty point three meters and determined the driver was at fault. 

\begin{quote}
\emph{“The x is equal to thirty one point zero five eight meters which is shorter than the skid mark length of forty point three meters. So yes the driver was at fault [R1].”}
\end{quote}

We note that Catherine got the answer right following the expected line of a reasoning. Her response pattern corresponds to {A3, A2, C2, C1, E2, R2, R3, A3, A2, C1, E2, and R1}. Our coding of Catherine's solution includes at least one code for each element in the ACER framework, indicating that the task met the minimum condition to elicit expected evidence to make conclusion about students' abilities to meet the LP. In addition to Catherine, if the task elicited expected evidence for majority of the students ($>50\%$ as we mentioned in Sec.~\ref{Analysis}), we concluded that the task can elicit students' abilities to meet the LP. Otherwise, it's required that we modify assessment tasks until they elicit the expected evidence to argue about students' proficiency to meet the LP.  

On the other hand, Catherine's written solution is associated with the response pattern, {A2, C1, E2, A2, C1, E2, R1}. The corresponding Written solution mirrors Catherine's problem-solving approach except for the subcodes A3, C2, R2, and R3, eliciting at least single evidence for each component of the ACER framework.

We applied the same process to all assessment tasks by taking into account the students' responses to those tasks to evaluate the tasks' potential to elicit the expected evidence in achieving the LP. If tasks ensured their potential to elicit expected evidence, we further explored the extent to which students' written solution mirrored that potential accurately. In the next section, we attend to some of the interesting aspects of our analysis with more details.

\subsubsection{Inter-rater Reliability}

%We first needed to ensure that the codebook is an accurate representation of student work. After the first author coded 53 instances of each transcribed verbal and written responses, the subset (5 instances for each transcribed verbal and written responses) were independently coded by another researcher. These 5 instances included different assessment tasks among multiple students in our data set. Afterwards, we discussed our coding to determine the utility of our codebook. The first round of discussion led us to do minor changes in the codebook. For example, we initially had another C code labeled as “C2- Distinguish the target parameter and the known parameters” which is no longer present. We merged this with “C1- Apply the general equations to a particular situation” as our discussion concluded that these two codes infer the same meaning of applying activated physics equations to a particular problem context. The other subcodes were then relabeled as they appear in the Appendix. Our changes also included refining wordings of the definitions of the subcodes such that it is understandable to a broad audience. 

After finalizing the codebook, 5 instances of transcribed verbal and written responses were independently coded by another researcher. These 5 instances included different assessment tasks among multiple students in our data set. %Prior to our discussion, we came to a 82\%, 67\%, 76\%, 63\%, and 55\% agreement on these 5 instances. 
After discussion, the coders came to a 100\% agreement on 4 of the 5 instances and a 96\% agreement on the remaining instance.% (that is the one with 67\% agreement prior discussion).

%After the second round of discussion, we came to a 100\% agreement on 4 of the 5 instances and a 96\% agreement on the remaining instance.

\subsubsection{Limitations}

One limitation of this work is associated with the assumption of the Think-Aloud protocol: verbalized information is the information acquired and heeded by the subject at a given time~\cite{Fonteyn1993}. However, the human thought processes are rapid enough such that it is likely that subjects verbalize a portion of their thoughts leaving other non-verbal. Thus, only problem-solving processes and verbalized reasoning should be used to make inferences about subjects' abilities. We also note that our data set includes a small population of students. Expanding the data set to include more (and more diverse) students and incorporating their reasoning in response to assessment tasks would be needed to strengthen their validity for something similar to a standardized assessment. However, we only mean to show this work as a proof of concept.

Another limitation of this work is that the fine-grained ESs are unique to these assessment tasks, and cannot be generalizable across different assessment tasks that address additional concepts and scientific practices. However, given the methodology, one needs to develop their own codebook based on the student evidentiary knowledge representations in their population of students. It is likely that the additional subcodes might appear during that process, but we argue that it cannot be significantly different between the similar problem types we presented in this manuscript.

\section{\label{Results and Discussion}Results and Discussion}

\subsection{\label{subsection1}Assessment Tasks Elicited the Expected Evidence for Students' Abilities}

%In this work, we developed assessment tasks that assess students' abilities to blend scientific practices with physics concepts by adopting and coupling principles from ECD, the 3D-LAP, and the ACER framework. We gave these tasks to students in Think-Aloud interview settings to evaluate the tasks' potential to elicit the expected evidence.

As noted in Table~\ref{Table3}, majority of the tasks (i.e., 6 out of 7) elicited students' reasoning that enabled us to capture their evidence pertaining to each component in the ACER framework \footnote{Though this is the minimum condition required to ensure the tasks' potential to elicit the expected evidence, it is typical that many students in our data set used numerous subcodes within each component in the ACER framework while meaningfully engaging in problem-solving.} except for the Ferris wheel task. % (i.e., 7/8, 7/8, 6/8, 4/7, 7/8, and 5/8 for each task respectively). 
This enabled us to capture that when students got the answers right, they got their answers for the right reasons, i.e, the expected cognitive processes were tapped. %Similarly, we inferred when students got the answers wrong, it was not because they misinterpreted the intent of the task.

\begin{table}[h]
\begin{ruledtabular}
\caption{The number of students who responded to each assessment task, the tasks that elicited the expected evidence, and number of students whose written solutions mirrored the elicited expected evidence are provided. ``N/A'' in the last column refers to the task (Task 3) that does not have the potential to elicit the expected evidence (i.e., Ferris wheel task).}
\centering
\begin{tabular}{r r r r} 
%\begin{tabular}{p{1.6cm}|p{1.4cm}|p{1.7cm}|p{2cm}}
%Assessment task & Number of students responded & Number of students Elicited the expected evidence & Written solutions mirrored the elicited expected evidence \\ \hline
Task \# & \# Responses & \# Evidence & \# Matched \\ \hline
1 & 7 & 7 & 5 \\ 
2 & 8 & 7 & 5 \\ 
3 & 7 & 1 & N/A \\
4 & 8 & 6 & 6 \\ 
5 & 7 & 4 & 4 \\ 
6 & 8 & 7 & 4 \\ 
7 & 8 & 5 & 4 \\ \hline
Total & 53 & 37 & 28 \\
\end{tabular}
\label{Table3}
\end{ruledtabular}
\end{table}

Situating our work in the ECD approach articulated in Harris \emph{et al.}~\cite{Harris2019} provides great insight into task design that assesses students' abilities to engage in scientific practices along with concepts. Our work strengthens the generalizability of ECD as a task design approach showcasing its potential to similarly extend into assessing student abilities to engage in scientific practices at introductory level physics courses. 

We also note that coupling the 3D-LAP with ECD to facilitate task features is promising when it comes to assessing students' abilities to blend scientific practices with concepts. Our work further validates the 3D-LAP as an effective tool to elicit student abilities to engage in scientific practices with the support of students' data. Thus, for task developers who have limited time, we suggest the 3D-LAP as a tool to begin with task development. However, we first recommend doing a thorough analysis of the domain of interest to determine the valued concepts to be assessed. This process can be followed by the integration of those concepts with the criteria for scientific practices of interest in the 3D-LAP to elicit student abilities to blend concepts with the scientific practices.

We note that utilizing the ACER framework to analyze both written and verbal work products takes student in-the-moment reasoning into account. Thus, this approach well-articulates the “framing” and “resources” perspectives that framework leverages. This work also expands the utility of the ACER framework to capture students' mathematical reasoning at introductory level.

Overall, we argue that a coherent, systematic approach to designing assessment tasks by coupling ECD with the 3D-LAP is productive when it comes to assessing students' abilities to blend scientific practices with concepts. Further, we argue that a framework that articulates what it means to use math in physics, i.e., the ACER framework guides our task validation process by capturing students' in-the-moment reasoning. 

\subsubsection{Modifying the task that failed to elicit `Using Math'}

As we explained above, 6 out 7 assessment tasks elicited the expected evidence which showcased students' abilities to achieve the LPs that address the scientific practice of ``Using Math". In this subsection, we explain how we can modify the remaining task that failed to elicit the expected evidence (i.e., Ferris wheel task) into a form which potentially would elicit the evidence as intended.

Unlike the assessment tasks that include numerical quantities in their problem statement (6 tasks), the task that includes symbolic variables, the Ferris wheel problem (1 task) did not prompt students to elicit the expected evidence. While our intention was that the Ferris wheel problem has the potential to elicit the expected evidence, 6 out of 7 students who responded started with a conceptual analysis of the problem to determine the positions where a rider in a Ferris wheel feels heaviest and the lightest. However, the follow-up question, “Approximately how large would $\omega$ have to be for this to have a noticeable effect on your weight?” prompted them to elicit the expected evidence. For example, given below is how William figured out in which positions the rider feels the heaviest and the lightest in the Ferris wheel problem. 

\begin{quote}
\emph{“So rotating counter-clockwise. Whenever it's moving up, the acceleration is kind of pulling it outwards so it's not really feeling like wait but when you're at the top hmm you're starting to go down the acceleration straight out so feels like you're moving up, so you're lightest at the top and going down. Heaviest at the bottom and going up just like an elevator.”}
\end{quote}

He started the problem doing a conceptual analysis of the problem as above, and then made an explanation to figure out the positions where the rider feels the heaviest and the lightest. Thereafter, answering the question about figuring out $\omega$ that might give a noticeable effect on weight, he showcased appropriate evidence as he figured out a reasonable expression for $\omega$. 

Thus, we believe that the structure of the variables in the form of symbolic or numerical might affect the way students activate their knowledge, skills, and abilities at hand. Though we see that students well-interpreted and elicited the expected evidence for the assessment tasks that include numerical variables, we do not see the same when it comes to the assessment task with symbolic variables. While we did not specifically probe the question during the interview about why students approached the way decided in response to the task that includes symbolic variables, this gives us some initial clues about the ways they interpret the tasks with respect to the nature of the variables in the problem statement. 

In particular, the utterances students made in response to the Ferris wheel task, \emph{“There is no numbers so…It seems kind of broad”, “[student is asking a question from the interviewer] So, with this question, how does it depend on diameter? Do you wanna leave those [symbolic variable of ``D" for diameter] in our answer?”} provided us initial evidence that they paid attention to the nature of the variables in the assessment tasks. However, more work is needed to strengthen the argument behind the dependability of the variable types in the problem statement that prompts students to elicit the expected evidence encouraging mathematical reasoning.

One potential future work is to intentionally design tasks that include both symbolic and numerical variables and explore their problem-solving approaches with respect to the variable types. Such work can give great insight into the ways in which we can prompt students to engage in more conceptual analysis of the problem rather than mere mathematical manipulations. Bringing our attention back to the task validation process, we further need to revise this task until it has the potential to elicit expected evidence to capture students' abilities to blend math with physics concepts.  

\subsection{Written Solutions Mirrored the Elicited Expected Evidence for Students' Abilities}

In order to address our third research question, we analyzed the extent to which the written solutions accurately represented the students' reasoning during problem-solving. For this analysis, we looked only at the six tasks that successfully elicited evidence of Using Math.  From those six tasks, 36 of the combined-codes included all four components of the ACER framework. In 28 out of those 36 instances, the written-codes covered the same elements of the ACER framework as the combined-codes (see Table~\ref{Table3}). In other words, though in these instances students elicited evidence for each component in the ACER framework, the reflection component (i.e., R1, R2, and R3, provided in Table~\ref{Table2}) is not mirrored in the written solution. %Also, the lack of evidence for the reflection components spanned across four assessment tasks suggesting modifications such that they have the potential to better elicit student reflections in the student written solution.

Students who engaged in reflections verbally, but did not include it in their written work might be an important aspect to further look into. This is because students' reflections during physics problem-solving are crucial and cognitively demanding. What students wrote down as part of working through a problem might be the things they believe instructors are valuing in their work. Therefore, by strengthening the importance of reflecting on responses students obtain to make sense of them at an earlier stage such as an introductory level is crucial. The lack of evidence for students' reflections in the written work suggests that we can modify our task features in a way that those elements are more conspicuous. In particular, we can scaffold the task prompting students to demonstrate proficiencies associated with reflections at an earlier stage. As students progress through the curriculum, the scaffolding can be removed to promote their autonomy to engage in reflection.

For example, Kang \emph{et al.}~\cite{Kang2014} show that high quality scaffolding can provide students opportunities to demonstrate their disciplinary proficiencies. Careful scaffolding to elicit student reasoning in assessments is also encouraged in the work by Cooper and Stowe~\cite{Cooper2018}. However, the scaffolding should not guide any specific problem-solving patterns. Rather the assessment tasks should allow students to construct solutions on their own to preserve the authenticity of the scientific practices along with concepts~\cite{NRC2001}. For example, we can explicitly guide students to utilize a self-constructed representation of a system that models a real-world phenomenon by including a question prompt similar to, ``construct a representation that models the physical system as part of your solution''. 

\section{\label{Conclusion}Conclusion}

In responding to the need for assessments to evaluate the extent course transformations are effective in addressing scientific practices, we demonstrated a principled task design approach that can be utilized to design tasks that assess student abilities to blend physics concepts (``force'' in our case) with the scientific practice ``Using Math''. As part of this process, we adopted ECD, and coupled that with the 3D-LAP to design assessment tasks (see Table~\ref{Table1}). 

We then used the ACER framework as a lens to look into students' responses to articulate the developed tasks' potential to elicit students' abilities to reasoning through mathematics when presented in Think-Aloud interview settings. This validation process takes into account both students' verbal responses and written solutions to holistically capture students' approaches to the presented assessment tasks. We updated the pre-defined ESs to accommodate student own knowledge representations emerging from the student data. The explicit validation process that includes written solutions expands our understanding about how these tasks can be modified for them to be utilized in paper-based summative assessment settings at large-scale college classrooms. Particularly in those settings, students' written solutions are the sole source to infer about the extent to which their learning progresses.  

Additionally, we explored the extent to which the written solutions accurately provide evidence of student reasoning. In addition to tasks' potential to elicit the expected evidence most of the time, students' written solutions too mirrored student reasoning most of the time.  

Therefore, we argue that utilizing and coupling both ECD and 3D-LAP is a productive approach to assess students' abilities to blend scientific practices with concepts. We also argue that a framework that articulates what it means to use math in physics, i.e., the ACER framework guides our task validation process by capturing students' in the moment reasoning. We note that Written solutions are reasonable artifacts to infer students' abilities to blend scientific practices with concepts. 

This work has important implications for PER-developed assessments. In particular, the approach to assessment task development adopting ECD and coupling with the 3D-LAP is promising at the introductory-level. Articulating an assessment argument ECD advocates which consists of the targeted performance, required knowledge, skills, and abilities to achieve the targeted performance, and evidence that supports students have the required knowledge, skills, abilities is crucial prior developing assessment tasks. The task features to elicit the determined evidence are informed by the the 3D-LAP. 

We validate the developed assessment tasks incorporating deeper insights into students' in the moment reasoning utilizing students' responses to these assessment tasks in Think-Aloud interview settings. Adopting an analytic framework -- the ACER -- helps us define what it refers to do math in physics, which is the target scientific practice for our study. Additionally, using the ACER framework and its perspectives on students' use of math minimizes the biases when analyzing data, in particular our own biases of what it refers to do math in physics.

In validating assessment tasks, we also placed emphasis on the students' written work, which is the sole source of information available for instructors to infer students' knowledge, skills, and abilities. This addition provides us insights into the ways in which we can modify assessment tasks such that the students' engagement in a task can be meaningfully elicited and inferred from their written work. Though for this work we only use students' abilities to blend ``math'' with ``force'', other assessment designers can use what they value in students' work and still follow the process articulated in this work.                 

In addition to the scientific practice of ``Using Math", we plan to expand this work to incorporate other scientific practices into our task design process. This future work will inform us about the extent to which our task design and validation process is consistent across different scientific practices. In the future, we plan to pilot our exam to a student population with multiple backgrounds to explore how these assessment tasks promote equity in a way that the factors like reading disabilities, for example, will not affect the student abilities to meet the expected learning performances. 

This work also informs our on-going work of developing a new standardized assessment for upper-division thermal physics -- The Thermal and Statistical Physics Assessment (TaSPA). In particular, the work presented in this paper informs the assessment tasks and the associated feedback for instructors in-development based on students' responses to these tasks~\cite{Jambuge2020}.

\begin{acknowledgments}
We are grateful for the students who volunteered for this study. We are thankful for the instructors in the classes in which we recruited students, for allowing us to make announcements about recruiting students for this study. We thank Lydia Bender for helping us recruiting students for this study. We are thankful for Bahar Modir for her kindness in sharing the interview tips and helping in finding video recording equipment. We appreciate Peter Nelson for his generosity for sharing video recording equipment when we requested. We appreciate Kim Coy's help in managing gift cards and promptly providing us them. We are thankful for Chris Hass for spending his time for our pilot interview. We thank Dean Zollman and Hien Khong for always being there for us to engage in wonderful discussions related to this project. We thank Katherine Ventura for designing, administering, and recording student responses to two assessment tasks that we adopt for our analysis. We are grateful for Paul Bergeron for participating in the Inter-rater reliability for this project and providing valuable feedback on this work. We thank Amogh Sirnoorkar for his valuable feedback on this work. We thank Department of Physics at Kansas State University for supporting this work. This material is based upon work supported by the National Science Foundation under Grant No. 1726360.  

\end{acknowledgments}

%\nocite{*}

\bibliography{Refs}% Produces the bibliography via BibTeX.

\appendix*

\section{Full Codebook}

% \usepackage{multirow}

\begin{table*}[h]
\centering
\caption{Full codebook with examples from data.}
\begin{tabular}{p{2cm}|p{2.5cm}|p{1cm}|p{6.25cm}|p{6.25cm}}
\hline\hline
Codes & Subcodes & Symbol & Examples from data (Verbal) & Examples from data (Written) \\ 
\hline
\multirow{3}{2 cm}{\shortstack[l]{Activation\\ (A)}}  & Identify appropriate physics concepts that can be used to understand the phenomenon & A1 & ``You feel `weight' as your net force/accleration'' & \(F_{\text{net}} = m\frac{D}{2}\omega^2 - mg \) \\ 
 \cline{2-5} 
 & Identify general physics equations to be applied & A2 & ``To   find acceleration, V squared equals V note squared plus two a hmm change in x component'' & \( V^2 = V_0^2 + 2a(x-x_0) \) \\ 
 \cline{2-5} 
& Identify target parameters & A3 & ``So then just find the initial speed and compare to see if the driver is at fault'' & \( V@D \)\\
\hline
 \multirow{4}{2 cm}{\shortstack[l]{Construction\\ of the model\\ (C)}} & Apply the general equations to a particular situation & C1 & ``Zero   is equal to two a x minus x note plus V note squared'' & \( 0 = V_0^2 + 2a(x-x_0) \) \\ 
 \cline{2-5} 
 & Make assumptions & C2 & ``I'm assuming there's no friction between rest to E. No friction that kind of   including drag. And energy is conserved and it will be sufficient. We're assuming that the train is attached to the track starting from rest'' & I assumed the track is frictionless \\ 
 \cline{2-5} 
 & Develop representations (diagrams, free body diagrams) & C3 & [Drawing   a free body diagram] ``You got force of friction, mg down hmm and then you got a velocity'' & [Free body diagrams/representations of the physical system/modified diagrams given in the exam] \\ 
 \cline{2-5} 
& Develop mathematical relations based on the physics concepts used & C4 & ``Thirteen [Fifteen] point two meters per second over twenty two point two [four] meters equals eighteen meters over x'' & \( {\frac{15.2 \;\text{m/s}}{22.4\;\text{m}} = \frac{18\;\text{m/s}}{X}} = 26.5\;\text{m} \) \\
\hline
 \multirow{5}{2 cm}{\shortstack[l]{Execution of\\ the\\ mathematics\\ (E)}} & Manipulate symbols & E1 & ``So I'm gonna use the Newton's law where F equals ma so a equals F over m'' & \( F = ma\),\newline \(a = \frac{F}{m} \) \\ 
 \cline{2-5} 
 & Perform an arithmetic calculation & E2 & [Input values to the calculator to calculate the values numerically] & \( V_0^2 = 415.65\),\newline \(V_0 = 20.397\)~m/s \\ 
 \cline{2-5} 
 & Execute math conceptually & E3 & ``m is just the same thing so m is cancelled out so a equals mu, k times g'' & \( ma = \mu_k mg,\) \newline \(a = \mu_kg \) \\ 
 \cline{2-5} 
 & Substitute expressions & E4 & ``Ok so, F equals ma which equals mu, k, m [g]'' & \( F = ma = \mu_k mg \) \\ 
 \cline{2-5} 
 & Manipulate mathematical expressions & E5 & ``Ok, we want to track off one thousand, nine point eight times ninety five meters and multiply all by two and divided by one thousand to get V, D squared'' & \( 0 = 15.2^2 + 2a(22.4),\)\newline \( \frac{-231.04}{2*22.4} = a \) \\ 
\hline
 \multirow{5}{2 cm}{\shortstack[l]{Reflection of \\the results \\(R)}} & Make sense of the answer with the information given in the prompt & R1 & ``I mean the average skid mark at this point would be twenty six point five meters and given forty point three which is like insane'' & \( X = 31.088\)~m \( < \;40.3\)~m, Yes the driver was at fault \\ 
 \cline{2-5} 
 & Make sense of the answer found in an intermediate/final step & R2 & ``The acceleration is equal to negative five point two one six which once again is a reasonable answer since they are going from a faster speed down to a lower speed'' & \( F = ma,\) \newline \(a = \frac{F}{m},\) \newline \(a = \frac{6749\;\text{N}}{1000\;\text{kg}},\) \newline \(a = -6.749\)~m/s$^2$ \\ 
 \cline{2-5} 
 & Make sense of the result for use in a subsequent step & R3 & ``Hmm, if that's the acceleration then that should also be the acceleration of the crash actually occurred on'' & [Calculated during previous part of the questions] \( a = -5.16\)~m/s$^2$,\newline \(V^2=V_0^2 + 2a(X-X_0),\)\newline \(\sqrt{V^2} = \sqrt{2*5.16*40.3} \)                                                 \\ \hline %originally -3
\hline
\end{tabular}
\label{Table4}
\end{table*}

\end{document}